\documentclass[aps,showpacs,onecolumn,floatfix,amsmath,amssymb,nofootinbib]{revtex4-1}

\usepackage[utf8]{inputenc}
\usepackage[english]{babel}%
\usepackage{cancel}
\usepackage{dcolumn}
\usepackage{tikz}
\usepackage{bm}%

\usepackage{amssymb}
\usepackage{amsmath}
\usepackage{epsfig}
\usepackage{graphicx}
\usepackage{color}
\usepackage{latexsym}
\usepackage{bm,latexsym}
\usepackage{mathrsfs}
\usepackage{float}
\usepackage{pbox}

\begin{document}

\title{(2+1)-dimensional Static Cyclic Symmetric Traversable Wormhole: Quasinormal Modes and  Causality. }

\author{Pedro Ca\~nate$^{1}$}
\email{pcanate@fis.cinvestav.mx, pcannate@gmail.com}

\author{Nora Breton$^{1}$}
\email{nora@fis.cinvestav.mx}

\author{Leonardo Ortiz$^{1}$}
\email{lortiz@fis.cinvestav.mx}

\affiliation{$^{1}$Departamento de F\'isica, Centro de Investigaci\'on y de Estudios Avanzados
del I.P.N.,\\ Apdo. 14-740, Mexico City, Mexico.\\}

\begin{abstract} 
In this paper we study a static cyclic symmetric traversable wormhole in $(2+1)-$dimensional  gravity coupled to  nonlinear electrodynamics in anti-de Sitter spacetime. The solution is  characterized by three parameters: mass $M$, cosmological constant $\Lambda$ and one electromagnetic parameter, $q_{\alpha}$. The causality of this spacetime is studied, determining its maximal extension and constructing then the corresponding  Kruskal-Szekeres and Penrose diagrams.
The quasinormal modes (QNMs) that result from considering  a massive scalar test field in the wormhole background are determined by solving in exact form the Klein-Gordon equation;  the  effective potential  resembles the one of a  harmonic oscillator shifted from its equilibrium position and, consequently,  the QNMs have a pure point spectrum.
\end{abstract}


\maketitle

\section{Introduction} 

Anti-de Sitter gravity in $(2+1)$-dimensions has attracted a lot of attention due to its connection to a Yang-Mills theory with the Chern-Simons term \cite{Witten2007}, \cite{Achucarro1986}.
Moreover, taking advantage of simplifications due to the dimensional reduction, three dimensional Einstein theory of gravity has turned out a good model from which extract relevant insights regarding the quantum nature of gravity \cite{Carlip1998}.  In three spacetime dimensions, general relativity becomes a topological field theory without propagating degrees of freedom. Additionally, in string theory, there are near extremal black holes (BHs) whose entropy can be calculated and have a near-horizon geometry containing the Ba\~nados-Teitelboim-Zanelli (BTZ) solution  \cite{BTZ1992}, \cite{BTZ1993}. Particularly for the (2+1)-dimensional BTZ black hole (BH), the two-dimensional conformal description has by now well established \cite{Carlip2005}: the BTZ-BH provides a precise mathematical model of a holographic manifold. For these reasons systems where the conformal description can be carried out all the way through are very valuable.

On the other hand nonlinear electrodynamics (NLED) has  gained interest for a number of reasons. Nonlinear electrodynamics consists of theories derived from Lagrangians that depend  arbitrarily on the two electromagnetic invariants, $F= 2(E^2-B^2)$ and $G= E \cdot B$, i.e. $L(F,G)$.  The ways in which $L(F,G)$ may be chosen are many, but two of them are outstanding: the Euler-Heisenberg theory  \cite{Heisenberg_36},  derived from quantum electrodynamics assumptions, takes into account some nonlinear features like the interaction of light by light.  And the Born-Infeld theory  \cite{BI}, \cite{Pleban}, proposed originally with the aim of avoiding the singularity in the electric field and the self-energy due to a point charge, it is a  classical effective theory that describes nonlinear features arising from the interaction of very strong electromagnetic fields,  where Maxwell linear superposition principle is not valid anymore.
Interesting solutions have been derived from the Einstein gravity coupled to NLED, like regular BHs, wormholes (WHs) sustained with NLED, among others, see for instance  \cite{Bronnikov2017}. It is also worth to mention that some NLED arise from the spontaneous Lorentz symmetry breaking (LSB), triggered by a non-zero vacuum expectation value of the field strength \cite{Urrutia2004}.

WHs in the anti- de Sitter (AdS) gravity  are interesting objects to study. For instance, regarding the transmission of information through the throat, the  understanding of the details of the traversable wormhole (WH) and its quantum information implications would  shed light  on the lost information problem  \cite{Bak2018}.
The thermodynamics of a WH  and its trapped surfaces  was addressed in \cite{PGDiaz2009}, establishing that the accretion of phantom energy, considered as thermal radiation coming out from the WH, can significantly widen the radius of the throat. In \cite{Maldacena2004} it is shown that Euclidean geometries with two boundaries that are connected through the bulk are similar to WH in the sense that they connect two well understood asymptotic regions.
In \cite{Gao2017} it is constructed a WH via a double trace deformation. Alternatively, WH solutions are constructed by gluing two spacetimes at null hypersurfaces, \cite{Kim2004}  \cite{Maeda2009} . Contrasting this procedure, 
in a recent paper the authors derived exact solutions of the Einstein equations coupled to NLED that can be interpreted as WHs and for certain values of the parameters such solutions  become the BTZ-BH \cite{Pedro2018}. Which has become an excellent laboratory for studying quantum effects since the seminal paper \cite{or94}. And regarding LSB, it can be mentioned as well that
WH solutions have been derived in the context of the bumblebee gravity \cite{Ovgun1}, their QNMs have been studied in \cite{Oliveira}, and the corresponding gravitational lensing in \cite{Ovgun2}.

Moreover, WHs are related to BHs; BH and WH spacetimes are obtained by identifying points in (2+1)-dimensional AdS space by means of a discrete group of isometries, some of them resulting in non-eternal BHs with collapsing WH topologies \cite{Aminneborg1998}.

In this paper we present an exact solution of the Einstein equations in (2+1)-dimensions with a negative cosmological constant (AdS) coupled to NLED. The solution can be interpreted as a WH sourced by the NLED field with a Lagrangian of the form $F^{1/2}$. This solution is a particular case of a broader family of solutions  previously presented in \cite{Pedro2018}.  The solution is characterized by three parameters: mass $M$, cosmological constant $- \Lambda=1 /l^2$ and the electromagnetic parameter $q_{\alpha}$.  The analogue to the Kruskal-Szekeres diagram  is constructed for the WH,  and the causality is investigated by means of the Penrose diagram, showing that the light trajectories traverse the WH.  The WH Penrose diagram resembles the anti-de Sitter one with the WH embedded in it.

A massive scalar test field is considered  in the WH background;  the corresponding Klein-Gordon (KG) equation, when written in terms of the tortoise coordinate,  acquires a Schr\"odinger-like form and it is solved in exact form determining the frequencies of the massive scalar field; the boundary conditions are  of purely ingoing waves at the throat and zero  outgoing waves at infinity.  The effective potential in the KG equation is a confining one and, accordingly,
we found that the spectrum is  real, showing then that the wormhole does not swallow the field as a black hole would, but the field goes through the throat passing then to the continuation of the WH, and preserving the energy of the test field. This also shows the stability of the scalar field in this WH background.

The outline of the paper is as follows. In the next section we present the metric for the WH and the field that sources it  as well as a brief review on its derivation. In Section III  we find the maximal extension and then the Penrose diagram is constructed. In Section IV the KG equation for a massive scalar field is considered in  the WH background. The  radial sector of the KG equation, when written in terms of the tortoise coordinate,  takes the form of a Schr\"odinger equation that is exactly solved, obtaining the QNMs by imposing the appropriate WH boundary  conditions. 
Final remarks are given in the last section. Details on the derivation of the QNMs as well as the setting of the boundary conditions are presented as an Appendix.


\section{The wormhole sourced by nonlinear electrodynamics}

The action of the (2+1) Einstein theory with cosmological constant, coupled to NLED  is given by

\begin{equation}\label{actionf}
S[g_{ab},A_{a}] 
= \int d^{3}x \sqrt{-g} \left( \frac{1}{16\pi}(R - 2\Lambda) + L(F)  \right),
\end{equation} 
where $R$ is the Ricci scalar and $\Lambda$ is the cosmological constant;  $L(F)$ is the NLED characteristic Lagrangian.
Varying this action with respect to gravitational field gives the Einstein equations,

\begin{equation}\label{EinsteinEqs}
G_{ab} + \Lambda g_{ab} = 8\pi E_{ab},
\end{equation}
where $E_{ab}$ is the electromagnetic energy-momentum tensor, 

 \begin{equation}
4\pi E_{ab} = g_{ab}L(F) - f_{ac}f_{b}{}^{c}L_{F}, \label{Eab}
\end{equation}
where $L_{F}$ stands for the derivative of $L(F)$ with respect to $F$ and  $f_{a b}$ are the components of the electromagnetic field tensor. 
The variation with respect to the electromagnetic potential $A_{a}$ entering in $f_{ab} = 2\partial_{[a} A_{b]}$, yields the electromagnetic field equations,


\begin{equation}\label{emEqs}
\nabla_{a}(L_{F}f^{ab}) = 0 = \nabla_{a}( _{\ast}\boldsymbol{f} )^{a},
\end{equation}
where $( _{\ast}\boldsymbol{f})^{a}$ is the dual electromagnetic field tensor which, for (2+1)-dimensional gravity, in terms of $f^{ab}$,  is defined by $(_{\ast} \boldsymbol{f})_{a} = \frac{\sqrt{-g}}{3}  \left( f^{tr} \delta^{\phi}_{a}  + f^{r\phi} \delta^{t}_{a}  + f^{\phi t} \delta^{r}_{a}   \right)$  with  $(a= t, r, \phi).$ 
We shall consider the particular nonlinear Lagrangian, $L(F)= \sqrt{-sF};$   these kind of Lagrangians have been called Einstein-power-Maxwell theories   \cite{Hassaine2008}, \cite{Gurtug2012}. On the other hand, in  \cite{Pedro2018} was shown that in $(2+1)$ Einstein theory coupled to NLED the most general form of the electromagnetic fields for stationary cyclic symmetric $(2+1)$ spacetimes, i.e., the general solution to Eqs. (\ref{emEqs}), is given by 
$_{\ast}\boldsymbol{f} = (g_{rr}c/\sqrt{-g})dr +  (a/3L_{F})dt + (b/3L_{F})d\phi$, where  $a$, $b$ and $c$ are constant, that by virtue of the Ricci circularity conditions, are subjected to the restriction that $ac=0=bc$. Therefore, in this geometry,  in order to describe the electromagnetic field tensor, we have two disjoint branches; [$a=0=b$, $c\neq0$] and [($a\neq0 \lor b\neq0$), $c=0$]. Here we are considering the branch $c\neq0$, and thus the only non-null electromagnetic field tensor component and the electromagnetic invariant are given, respectively,  by 
\begin{equation}\label{Finvar}
f^{\phi t}=\frac{3 g_{rr} c}{(\sqrt{-g})^2}, \quad F = \frac{1}{2}f^{\phi t}f_{\phi t} =  \frac{9}{2}\frac{c^{2}}{g_{tt}g_{\phi\phi}}. 
\end{equation}
With these assumptions a five-parameter family of solutions with a charged rotating wormhole interpretation was previously presented in \cite{Pedro2018}. In this work we shall address in detail the (2+1)-dimensional static cyclic symmetric wormhole. 

For the sake of completeness, we give a brief review on the derivation of the solution.
The field equations of general relativity (with cosmological constant) coupled to NLED for a static cyclic symmetric (2+1)-dimensional spacetime with line element

\begin{equation}\label{ansatz}
ds^{2} =  -  N^{2}(r) dt^{2} + \frac{dr^{2}}{f^{2}(r)}  + r^{2} d\phi^{2},
\end{equation}
written in the orthonormal frame $\{$ $\theta^{(0)} = N(r) dt$, $\theta^{(1)} = \frac{dr}{f(r)}$, $\theta^{(2)} = r d\phi$ $\}$, are given by 

\begin{eqnarray}
&& G_{(0)}{}^{(0)} = 8\pi E_{(0)}{}^{(0)} - \Lambda \delta_{(0)}^{(0)}  \quad  \Rightarrow  \quad  \frac{ (f^{2})_{,r} }{2r} = 2 \left( L - 2FL_{F}  \right) - \Lambda, \label{field_C_1}\\
&&G_{(1)}{}^{(1)} = 8\pi E_{(1)}{}^{(1)} - \Lambda \delta_{(1)}^{(1)}  \quad \Rightarrow  \quad  \frac{ f^{2} N_{,r} }{rN} = 2L - \Lambda, \label{field_C_2}\\
&&G_{(2)}{}^{(2)} = 8\pi E_{(2)}{}^{(2)} - \Lambda \delta_{(2)}^{(2)}  \quad \Rightarrow  \quad \frac{ f (f N_{,r} )_{,r} }{N}  = 2 \left(  L - 2FL_{F}  \right) - \Lambda, \label{field_C_3}
\end{eqnarray}
where the comma denotes ordinary derivative with respect to the radial
coordinate $r$. 

The metric given by

\begin{equation}\label{NewSolution1}
ds^{2} =  -  \left( - q_{\alpha}Mr    +  q_{\beta} \sqrt{ - M   - \Lambda r^{2}}  \right)^{2} dt^{2} + \frac{dr^{2}}{ - M  - \Lambda r^{2} }  + r^{2} d\phi^{2}, 
\end{equation}  
is a solution of the Einstein-NLED field equations, with cosmological constant, with the nonlinear electromagnetic Lagrangian $L(F)= \sqrt{-sF}$, whose electromagnetic field tensor is given by (\ref{Finvar}) and with the electromagnetic parameter $c$ given by $c = \sqrt{2}M^{2}  q_{\alpha}/ (6 \sqrt{s}) $.

In order to obtain the solution (\ref{NewSolution1}), note that  $L(F)= \sqrt{-sF}$ is such that  $\left( L - 2FL_{F}  \right) = 0$, then Eq.(\ref{field_C_1}) becomes, 

\begin{equation}\label{f2dr} 
 (f^{2})_{,r} = - 2\Lambda r \Rightarrow  f^{2}(r) = - M  - \Lambda r^{2},  
\end{equation}
with  $M$ being an integration constant.
On the other hand, according to (\ref{Finvar}), for the line element (\ref{ansatz}) the invariant $F$ takes the form, 
 
\begin{equation}\label{FNa}
F = -\frac{1}{2} \left( \frac{3c}{rN}  \right)^{2}.
\end{equation}
If one replaces $F$ from Eq.(\ref{FNa}) into $L(F)$ in Eq. (\ref{field_C_2}), we arrive at 

\begin{equation}
\frac{ f^{2} N_{,r} }{rN} = 2L - \Lambda  \Rightarrow \frac{ \left(- M  - \Lambda r^{2} \right) N_{,r} }{rN} 
= 2\sqrt{\frac{s}{2} \left( \frac{3c}{rN}  \right)^{2} } - \Lambda  \Rightarrow \left(- M  - \Lambda r^{2} \right) N_{,r}  + \Lambda r N = 3\sqrt{2s}c.   
\end{equation} 
Now, by substituting $c = \sqrt{2}M^{2}  q_{\alpha}/ (6 \sqrt{s}) $ into the previous equation, yields

\begin{equation}
\left(- M  - \Lambda r^{2} \right) N_{,r}  + \Lambda r N  =  M^{2}  q_{\alpha},  
\end{equation} 
whose  general solution is

\begin{equation}\label{Ndr}
N (r) = - q_{\alpha}Mr  +  q_{\beta} \sqrt{ - M   - \Lambda r^{2}}, 
\end{equation}
where $q_{\beta}$ is an integration constant. 
Finally, by substituting (\ref{f2dr}) and (\ref{Ndr}) into (\ref{field_C_3}), one finds  

\begin{equation}
\frac{ f (f N_{,r} )_{,r} }{N}   = \sqrt{ - M  - \Lambda r^{2} } \frac{  \left( q_{\alpha}M \frac{\Lambda r}{\sqrt{- M  - \Lambda r^{2}}} -  q_{\beta}\Lambda  \right)  }{ - q_{\alpha}Mr  +  q_{\beta} \sqrt{ - M   - \Lambda r^{2}} } =  \frac{  \left( q_{\alpha}M r -  q_{\beta} \sqrt{ - M  - \Lambda r^{2} } \right) \Lambda }{ - q_{\alpha}Mr  +  q_{\beta} \sqrt{ - M   - \Lambda r^{2}} } =  - \Lambda,
\end{equation}
such that Eq. (\ref{field_C_3}) is trivially satisfied by the Lagrangian $L = \sqrt{-sF}$, the structural functions $f^{2}(r)$, $N^{2}(r)$ given by (\ref{f2dr} ) and (\ref{Ndr}), and the electromagnetic field given by (\ref{Finvar}).\\


\subsection{Wormhole properties}

Let us show that the solution (\ref{NewSolution1}) allows a  traversable wormhole interpretation. \\  
The canonical metric for a (2+1)-dimensional static cyclic symmetric WH \cite{ThorneMorris} is given by

\begin{equation}\label{WH_Thorne}
ds^{2} =  -  e^{2 \Phi(r)} dt^{2} + \frac{dr^{2}}{ 1  - \frac{b(r)}{r} }  + r^{2} d\phi^{2}.
\end{equation}
By comparison with (\ref{NewSolution1}) we see that $e^{\Phi(r)} = -q_{\alpha}Mr +q_{\beta} \sqrt{ - M   - \Lambda r^{2}}$ and $b(r)=r(1+M + \Lambda r^2)$, where $- \Lambda = 1/ l^2$. In this paper the case $q_{\beta}=0$ will be the subject of our study,  

\begin{equation}\label{NewSolution2}
ds^{2} =  -  \left( - q_{\alpha}Mr \right)^{2} dt^{2} + \frac{dr^{2}}{\frac{ r^{2}}{l^2} -M}  + r^{2} d\phi^{2} \quad \textup{  with   } \quad M > 0.
\end{equation}

Then we can check the  WH properties of the metric (\ref{NewSolution2}):

(i) The existence of a throat $r_0$ where $b(r_0)= r_0$. Such a throat is located at $r_0=   \sqrt{l^{2}M}$. The range of the $r$-coordinate is in the interval $r \in [r_0, \infty)$.

(ii) The absence of horizons. It  is fulfilled since  $ e^{2 \Phi(r)}=( - q_{\alpha}Mr)^{2}$ is nonzero for all  $r \in [r_0, \infty)$.   

(iii) The fulfilment of the flaring out condition that is related to the traversability of the WH.   We shall see that traversability has a consequence on the form of the QNMs. This  condition is guaranteed if
the derivative of $b(r)$ when evaluated at the throat is less than one, $b'(r_0) < 1$; in our case, $b'(r_0)= 1-2M < 1$.

The nonlinear field in our case is generated by the Lagrangian  $L(F)= \sqrt{-sF},$  where $F$,  the electromagnetic invariant, and the only non-vanishing electromagnetic component,  $f_{t \phi}$,  are given,  respectively,  by

\begin{equation}
F=- \frac{M^2}{4 s r^4}, \quad  f_{t \phi}= - \partial_{\phi} A_t= \frac{q_{\alpha} M^2}{\sqrt{2s}}.
\end{equation}

Moreover, it is well known that in GR matter obeying the standard energy conditions is not worth to open a throat and so create a traversable wormhole. In the case we are analyzing,  the NLED energy-momentum tensor does not satisfy  the null energy condition (NEC), rendering this into a traversable WH. To check the violation of  the NEC due to NLED,  let us consider the  null vector in the orthonormal frame, $ \boldsymbol{n} = (1,1,0), $  and calculate $ E_{(\alpha)(\beta)}n^{(\alpha)}n^{(\beta)} = E_{(0)(0)} + E_{(1)(1)} = L(F)/(4 \pi)$, then,  using (\ref{field_C_2}) to determine $L(F)$,  we obtain that  

\begin{equation}
 E_{(\alpha)(\beta)}n^{(\alpha)}n^{(\beta)}   
 = - \frac{M}{8 \pi r^2} < 0,  
\end{equation}
from which we see that NEC is violated; particularly, evaluating at the throat $r_0^2= M l^2$, $E_{(\alpha)(\beta)}n^{(\alpha)}n^{(\beta)}= - 1/(8 \pi l^2)$.


\section{The maximal extension and causality: Kruskal-Szekeres and Penrose diagrams }\label{Max_Ext}

In order to understand  the causal structure and the structure at infinity of the WH with metric  (\ref{NewSolution2}),
we will construct its Penrose diagram.  Following the standard  procedure we derive first the analogue to the Kruskal-Szekeres diagram.
 
To start with,  since the causal structure is defined by the light cones,  we need to consider the radial null curves which by definition satisfy 
the null condition $0 = ds^{2}(k^{\alpha},k^{\beta}) $, $k^{\alpha}$ being a null vector;  that implies 

\begin{equation}
\frac{dt}{dr} =  \pm \frac{1}{\sqrt{ ( r^{2}/l^2 -M)q_{\alpha}^{2}M^{2}r^{2} } }.
\end{equation}

Since the metric (\ref{NewSolution2}) has a coordinate singularity at $r = \sqrt{ -\frac{M}{\Lambda} } =  \sqrt{l^{2}M}$, we shall use the tortoise coordinate $r_{\ast}$
defined by 

\begin{equation}\label{tortuga}
\frac{dr_{\ast}}{dr} = \sqrt{-\frac{g^{tt}}{g^{rr}}} = \frac{1}{\sqrt{ ( {r^{2}}/{l^2} - M  ) \left(  q_{\alpha} Mr     \right)^{2} }}.    
\end{equation}

Integrating Eq. (\ref{tortuga})  for the tortoise coordinate,  $r_{\ast}$, we obtain

\begin{equation}\label{r_ast_wh}
r_{\ast}  =  -\frac{i}{ 2  \sqrt{ q_{\alpha}^{2} M^{3} } }\ln{\!\left(  \frac{ \sqrt{ M -{r^{2}}/{l^{2}}  } + \sqrt{M} }{ \sqrt{ M -{r^{2}}/{l^{2}}  } - \sqrt{M} } \right) }.
\end{equation}
We should remark that  $r_{\ast}$ is real, in spite of how it looks  Eq. (\ref{r_ast_wh}). It turns out that  $r_{\ast}$ in the previous form is very convenient when applying the WH boundary conditions to the KG equation.  It can be shown that $r_{\ast}$ can be written equivalently as 

\begin{equation}\label{r_ast_wh2}
r_{\ast}  =  - \frac{ 1 }{ \sqrt{ q_{\alpha}^{2} M^{3} } } \tan^{-1} \left({ \sqrt{\frac{M}{ {r^2}/{l^2}-M}}} \right).
\end{equation}

\begin{figure}
\centering
\includegraphics[width=17.5cm,height=6cm]{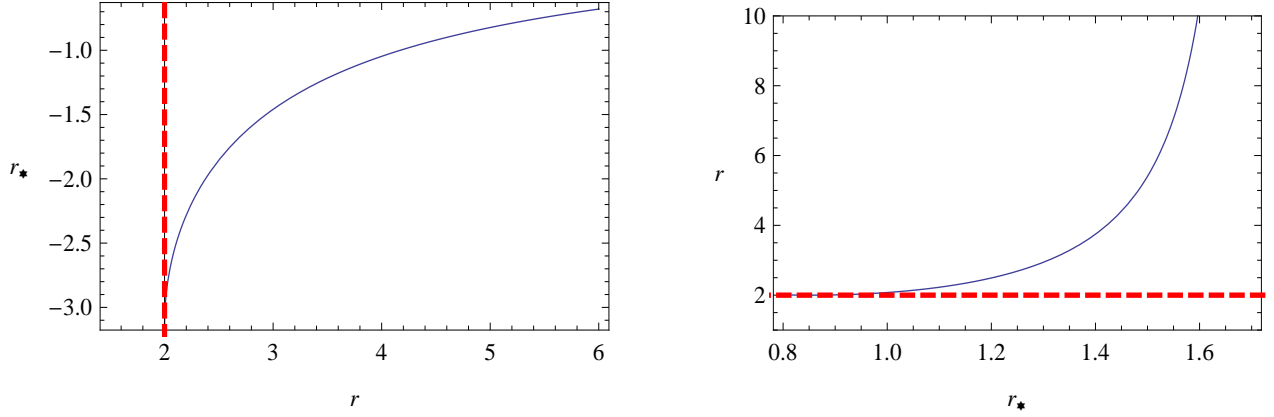}
\caption{\label{fig1} The tortoise coordinate $r_{\ast}$ as a function of the  coordinate $r$, Eq. (\ref{r_ast_wh2}), as well as its inverse, Eq. (\ref{r_rast}) (right) are plotted; the dashed red  straight lines show the position of the throat, $r_0^2=M l^2$. The parameters are fixed as $M=1$, $q_{\alpha}=0.5$ and the AdS parameter is $l=2$. }
\end{figure}

Since the function $\tan(x)$ is periodic, then $r_{\ast}$ is not uniquely defined in terms of $r$, i.e., for each value of $r$ there are multiple values of $r_{\ast}$, $r_{\ast} + \frac{1}{  \sqrt{ q_{\alpha}^{2} M^{3} } } \pi\xi_{_{n}}$, with $\xi_{_{n}} \in \mathbb{Z}$.  
The range of $r_{\ast}$  is determined by its values at the throat, $r_0$, and at infinity:  at the throat $ r_{\ast}(r_0) =
\frac{1}{  \sqrt{ q_{\alpha}^{2} M^{3} } }(-\frac{\pi}{2} + \pi\xi_{_{n}})$, while at the AdS infinity, $r \sim \infty$,  $r_{\ast} \sim \frac{\pi\xi_{_{n}}}{  \sqrt{ q_{\alpha}^{2} M^{3} } }$, where $\xi_{_{n}}$ is the integer defining each particular branch. Since all these branches are equivalent, we select the branch $\xi_{_{n}}=1$; consequently, the range of the tortoise coordinate is  $\frac{\pi}{2 \sqrt{ q_{\alpha}^{2} M^{3} } }  \leq r_{\ast} < \frac{\pi}{   \sqrt{ q_{\alpha}^{2} M^{3} } }$. 
From Eq. (\ref{r_ast_wh2}) we can obtain $r(r_{\ast})$,

\begin{equation}\label{r_rast}
r^2= M l^2 \left[{ 1 + \cot^2\!\left(  \sqrt{q_{\alpha}^{2} M^{3}} r_{\ast} \right) }\right]= M l^2 \csc^2\!\left( \sqrt{q_{\alpha}^{2} M^{3}} r_{\ast}\right).
\end{equation}
The tortoise coordinate as a function of $r$ as well as its inverse are shown in Fig. \ref{fig1}.

In terms of the coordinates $(t, r_{\ast}, \phi)$ the line element (\ref{NewSolution2}) becomes

\begin{equation}\label{metric_t_r_ast}
ds^{2}  = q_{\alpha}^{2}M^{2}r^{2}\left( - dt^{2} + dr^{2}_{\ast} \right) + r^{2}d\phi^{2}.
\end{equation}
In terms of these coordinates the radial null geodesics satisfy $t = \pm r_{\ast} +$ constant. This motivates us to define the advanced and retarded null coordinates $v$ and $u$,  respectively,  by 

\begin{equation}\label{null_u_v}
v = t + r_{\ast} \quad \textup{ and } \quad  u = t - r_{\ast},    
\end{equation}
where $-\infty < v < \infty$, $-\infty < u < \infty$. In these coordinates  the metric (\ref{metric_t_r_ast}) becomes

\begin{equation}\label{metric_u_v}
ds^{2}  = - q_{\alpha}^{2} M^{3} l^2 \csc^{2}{ \left( \frac{u - v}{2}\sqrt{q_{\alpha}^{2} M^{3} } \right) }     du dv + r^{2}d\phi^{2}.
\end{equation}

From Eq. (\ref{r_rast}) and using $r_{\ast} = (v-u)/2 $, it can be determined $r$ as a function of $(u,v)$ as, 

\begin{equation}\label{rde_v_u}
r^{2} = l^{2}M\csc^{2}{ \left( \frac{(v - u) }{2}\sqrt{ q_{\alpha}^{2} M^{3} } \right) }.    
\end{equation}

Despite  the coordinate ranges $-\infty < u < \infty$ and $-\infty < v < \infty$,  the metric (\ref{metric_u_v}) spans only on the region $\frac{\pi}{2 \sqrt{ q_{\alpha}^{2} M^{3} }} \leq r_{\ast} = \frac{v - u}{2} < \frac{\pi}{ \sqrt{q_{\alpha}^{2} M^{3} }}$.  In order to extend the spacetime beyond the wormhole throat,  $r_0 =  \sqrt{l^{2}M}$, we are going to determine the affine parameter, $\tau$,  along the null geodesics and reparametrize them with the  coordinates $V = V(v)$, and $U = U(u)$. 
We know that the geodesic tangent vector $K = K^{\beta}\partial_{\beta} = \frac{ dx^{\beta} }{ d \tau }\partial_{\beta}$ satisfies  

\begin{equation}\label{affineP}
 K^{\alpha} \nabla_{\alpha} K^{\beta} = 0. 
 \end{equation}

The tangent vector can be written as $K = \frac{ dr }{ d \tau }\left( \pm \frac{1}{\sqrt{ (r^{2}/l^2 -M)q_{\alpha}^{2}M^{2}r^{2} } } \partial_{t} + \partial_{r}  \right) $. Thus, by substituting $K^{\beta}$ into (\ref{affineP}), we find for the affine parameter $ \tau$,

\begin{equation}
\tau =  {C}_{0} \cot{ \left( \frac{u - v }{2}\sqrt{ q_{\alpha}^{2} M^{3} } \right) } +  C_{1}, 
\end{equation}
where ${C}_{0} $ and $C_{1}$ are integration constants. 
Then, the affine parameter along the null geodesics suggests to define the new coordinates $U$ and $V$ as

\begin{equation}\label{Udeu_Vdev}
U = \cot{ \left( \frac{u}{2}\sqrt{ q_{\alpha}^{2} M^{3} } \right) } \quad \textup{ and } \quad V = \cot{ \left( -\frac{v}{2}\sqrt{ q_{\alpha}^{2} M^{3} } \right), } \quad \textup{ with ranges } \quad -\infty < U, V < \infty ,
\end{equation}
therefore, in terms of $U$ and $V$ the metric (\ref{metric_u_v}) becomes 

\begin{equation}\label{metric_U_V}
ds^{2} =  \frac{4 l^2}{(U+V)^2} dU dV + r^2 d\phi^{2},
\end{equation}
where we have used that $r^2= l^2M (1 + U^{2})(1 + V^{2} )/(U+V)^2$.
By transforming to  $X = (V+U)/2$ and $T = (V-U)/2$, the metric (\ref{metric_U_V}) can be reduced to a more usual form given by   

\begin{equation}
ds^{2}  =  \frac{ l^{2} }{ X^{2} }  \left( -dT^{2} + dX^{2} \right) + r^2 d\phi^{2}. 
\end{equation}

We can see that the coordinates $(T,X)$ are the analogue to the Kruskal coordinates in Schwarzschild
spacetime. In terms of $(t,  r_{\ast})$ the coordinates $(T, X)$ are 

\begin{equation}\label{X_T_t_r}
X^2-T^2=UV= \frac{\cos \left({t \sqrt{q_{\alpha}^{2}M^3} }\right) + \cos \left({ r_{\ast} \sqrt{q_{\alpha}^{2}M^3}}\right)}{\cos \left({t \sqrt{q_{\alpha}^{2}M^3}}\right) - \cos \left({ r_{\ast} \sqrt{q_{\alpha}^{2}M^3}}\right)},  \quad \quad  \frac{ X + T }{X - T} = \frac{V}{U} =  \frac{ \sin{\left( r_{\ast}\sqrt{ q_{\alpha}^{2} M^{3} } \right)} - \sin{\left( t\sqrt{ q_{\alpha}^{2} M^{3} } \right)  } }{ \sin{\left( r_{\ast}\sqrt{q_{\alpha}^{2} M^{3} } \right)} + \sin{\left(  t\sqrt{q_{\alpha}^{2} M^{3} } \right) } }.
\end{equation}
from the above equations we deduce that, in terms of $T$ and $X$,  the region corresponding to the wormhole throat $r=r_{0}=  \sqrt{l^{2}M}$, or  $r_{\ast} = \frac{\pi}{2 \sqrt{ q_{\alpha}^{2} M^{3} } }$ is determined by the two equations
 
\begin{equation}\label{TX_garganta}
X^2-T^2 = 1 \quad \textup{ and } \quad  \frac{ X + T }{X - T} =  \frac{ 1 - \sin{\left( t\sqrt{ q_{\alpha}^{2} M^{3} } \right)  } }{ 1 + \sin{\left(  t\sqrt{q_{\alpha}^{2} M^{3} } \right) } } 
\end{equation}
Since $X^2-T^2 = 1$, then $X\neq T$ and  $\frac{ X + T }{X - T} =  \frac{ 1 - \sin{\left( t\sqrt{ q_{\alpha}^{2} M^{3} } \right)  } }{ 1 + \sin{\left(  t\sqrt{q_{\alpha}^{2} M^{3} } \right) } }$ can be written as 

\begin{equation}
T = m(t) X, \quad \textup{ with } \quad m(t) = -\sin(t) \in (-1,1),        
\end{equation}
$T = m(t) X$ corresponds to the region determined by all the straight lines that cross the origin $(X=0,T=0)$ with slope between $-1$ and $1$. The intersection between $X^2-T^2 = 1$ and $T = m(t) X$, yields $X^2-T^2 = 1$. Thus, the region corresponding to the wormhole throat in terms of $X$ and $T$, corresponds to $X^2-T^2 = 1$, i.e. the hyperbola with vertices at $(X=\pm1,T=0)$. 

From Eqs. (\ref{X_T_t_r}) can be obtained that

\begin{equation}\label{r_X_T}
 T^{2}  \sin^{2}{\left( r_{\ast}\sqrt{ q_{\alpha}^{2} M^{3} } \right)} + X^{2} \left( \frac{T^{2} - X^{2} - 1}{T^{2} - X^{2} + 1} \right)^{2} \cos^{2}{\left( r_{\ast}\sqrt{ q_{\alpha}^{2} M^{3} } \right)} = X^{2}.    
\end{equation}
Regarding infinity, from the previous equation,  the asymptotic AdS region $r\sim\infty$, or $r_{\ast} \sim \frac{\pi}{ \sqrt{ q_{\alpha}^{2} M^{3} } }$, in terms of $X$ and $T$,  is given by 

\begin{equation}\label{TX_r_infinito}
X^{2}\left( \frac{T^{2} - X^{2} - 1}{T^{2} - X^{2} + 1} \right)^{2} \sim X^{2},    
\end{equation}
which is fulfilled by $X \sim 0$, or by the region\footnote{  
For $r_{\ast}\sim \frac{\pi}{ \sqrt{ q_{\alpha}^{2} M^{3} } }$ and $t\in \mathbb{R}$, Eqs. (\ref{X_T_t_r}) become
\begin{equation}\label{AdS_X_T_t_r}
X^2-T^2= \frac{ \cos\left({t \sqrt{q_{\alpha}^{2}M^3} }\right) - 1^{-} }{ \cos\left({t \sqrt{q_{\alpha}^{2}M^{3}}}\right) + 1^{-}},  \quad \quad  \frac{ X + T }{X - T} = \frac{V}{U} =  \frac{ 0^{+} - \sin{\left( t\sqrt{ q_{\alpha}^{2} M^{3} } \right)  } }{ 0^{+} + \sin{\left(  t\sqrt{q_{\alpha}^{2} M^{3} } \right) } }, \quad\textup{ being}\quad 0 \lessapprox 0^{+}, \textup{ and } 1^{-} \lessapprox 1.
\end{equation}
\begin{itemize}
\item For $\frac{(2n+1)\pi}{ \sqrt{q_{\alpha}^{2}M^3}} \neq t \not\approx \frac{(2n+1)\pi}{ \sqrt{q_{\alpha}^{2}M^3}}$ 
with $n\in\mathbb{Z}$, the Eqs. (\ref{AdS_X_T_t_r}), yields $X=0$ and $T^2 = \frac{1-\cos \left({t \sqrt{q_{\alpha}^{2}M^3} }\right) }{1 + \cos\left({t \sqrt{q_{\alpha}^{2}M^3}}\right) }$.
\item For  $t = \frac{(2n+1)\pi}{ \sqrt{q_{\alpha}^{2}M^3}}$ with $n\in\mathbb{Z}$, the Eqs. (\ref{AdS_X_T_t_r}), yields $X^2 - T^2 \gg 1$ and $|X|\gg|T|$.
\item For  $t\neq \frac{(2n+1)\pi}{ \sqrt{q_{\alpha}^{2}M^3}}$ with $t\sim \frac{ (2n+1)\pi^{-} }{ \sqrt{q_{\alpha}^{2}M^3}}$, or $t\sim \frac{ (2n+1)\pi^{+} }{ \sqrt{q_{\alpha}^{2}M^3}}$, with $n\in\mathbb{Z}$, the Eqs. (\ref{AdS_X_T_t_r}), yields $T^2 - X^2 \gg 1$ and $|X|\ll|T|.$
\end{itemize}
}, 

\begin{equation}\label{TX_infinito}
\frac{T^{2} - X^{2} - 1}{T^{2} - X^{2} + 1} \sim 1   \Rightarrow \left\{ \quad (T^{2} - X^{2}) \gg 1 \quad \vee \quad (X^{2} - T^{2}) \gg 1 \quad \right\} 
\end{equation}

Collecting the regions defined by (\ref{TX_garganta}) and (\ref{TX_r_infinito}), the  Kruskal diagram corresponding to the spacetime (\ref{NewSolution2}) is depicted in Fig. \ref{anlKrusk}.

\begin{center}
\begin{figure}[htb!]
\begin{tikzpicture}[scale=1.6]
\draw[<->] (-2.3,0) -- (2.3,0) node[right] {$X$}; 
\draw[<->] (0,-2.5) -- (0,2.5) node[left] {$T$};
\draw[smooth, domain = 0.707107:2.2, color=blue]
plot (\x,{ ((\x)^2 - 0.5)^(1/2)  });
\draw[smooth, domain = 0.707107:2.2, color=blue]
plot (\x,{ -((\x)^2 - 0.5)^(1/2)  });
\draw[smooth, domain = -0.707107:-2.2, color=blue]
plot (\x,{ ((\x)^2 - 0.5)^(1/2)  });
\draw[smooth, domain = -0.707107:-2.2, color=blue]
plot (\x,{ -((\x)^2 - 0.5)^(1/2)  });
\draw[dashed, domain = -2.1:2.1, color=black]
plot (\x,{\x}); 
\draw[dashed, domain = -2.1:2.1, color=black]
plot (-\x,{\x}); 
\draw[smooth, domain = -1.9:1.9, color=red]
plot (\x,{ ((\x)^(2) + 3)^(1/2)  }); 
\draw[smooth, domain = -1.9:1.9, color=red]
plot (\x,{ -((\x)^(2) + 3)^(1/2)  }); 
\draw[smooth, domain = 1.73206:2.4, color=red]
plot (\x,{ ((\x)^2 - 3)^(1/2)  });
\draw[smooth, domain = 1.73206:2.4, color=red]
plot (\x,{ -((\x)^2 - 3)^(1/2)  });
\draw[smooth, domain = -1.73206:-2.4, color=red]
plot (\x,{ ((\x)^2 - 3)^(1/2)  });
\draw[smooth, domain = -1.73206:-2.4, color=red]
plot (\x,{ -((\x)^2 - 3)^(1/2)  });
\draw[<-, dashed] (-2.1,1.9) -- (-2.7,1.9) node[left] {$(X^{2}-T^{2}=1) \equiv (r=r_{0})$};
\draw[<-, dashed] (1.99,1.8) -- (2.6,1.8) node[right] {$(X^{2}-T^{2}=1) \equiv (r=r_{0})$};
\draw[<-, dashed] (-1.6,-2.3) -- (-2,-2.3) node[left] {$(T^{2} - X^{2} \gg 1) \equiv(T\sim-\infty)$};
\draw[<-, dashed] (1.6,2.3) -- (2,2.3) node[right] {$(T^{2} - X^{2} \gg 1) \equiv (T\sim\infty)$};
\draw[<-, dashed] (-2.182,-1.2) -- (-2.6,-1.2) node[left] {$(X^{2} - T^{2} \gg 1) \equiv(r\sim\infty)$};
\draw[<-, dashed] (2.2,-1.3) -- (2.6,-1.3) node[right] {$(X^{2} - T^{2} \gg 1) \equiv (r\sim\infty)$};
\draw[-,dashed,color=red] (-0.005,-2.5) -- (-0.005,2.5);
\draw[-,dashed,color=red] (0.005,-2.5) -- (0.005,2.5); 
\draw[<-, dashed] (0.04,2.1) -- (1.2,2.83) node[right] {$(X=0) \equiv (r\sim\infty)$};
\draw[-,dashed,color=black] (0.71,-0.02) -- (0.71,0.1);
\draw[-,dashed,color=black] (-0.72,-0.02) -- (-0.72,0.1);
\draw (0.71,0.01) node[below] {{\small $1$}};
\draw (-0.79,0.01) node[below] {{\small $-1$}};
\end{tikzpicture}
\caption{ It is shown the analogue to the Kruskal-Szekeres diagram for the WH. The blue hyperbola with vertices at $(X,T)=(\pm1 , 0)$, represents the WH throat. Horizontal hyperbolae with vertices at $(X,T)=(\pm\infty,0)$, as well as the region $X=0$, represent spatial infinity; the vertical hyperbola  (vertices at $(X,T)=(0, \pm\infty)$) represents time infinity.
In the Kruskal diagram, the hyperbola at the top (time infinity) can be identified with the one at the bottom, in case we want to work with the covering space.}
\label{anlKrusk}
\end{figure}
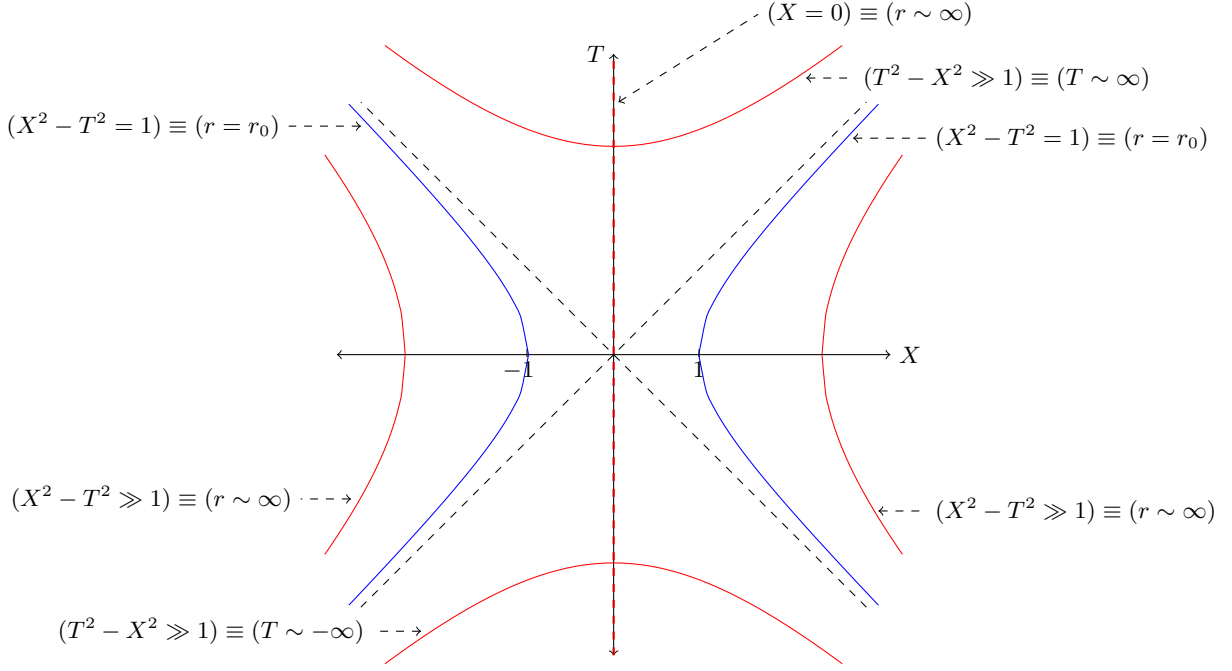
\end{center}
In order to obtain the Penrose diagram of the spacetime in consideration, we introduce the coordinates $\lambda$ and $\rho$ given by     

\begin{equation}
T - X = \tan{ \left( \frac{\lambda - \rho }{2} \right) },  \quad T + X = \tan{ \left( \frac{\lambda + \rho }{2} \right) }.
\end{equation}   

Then, the WH metric in terms of $(\lambda, \rho)$, yields

\begin{equation}\label{metric_l_r}
ds^{2}  =  l^2 \csc^{2}{\rho} \left( -d\lambda^{2} + d\rho^{2}  + M d\phi^{2} \right) =  \frac{l^2}{ \sin^{2}{\rho} } \left( -d\lambda^{2} + d\rho^{2} +  d\tilde{\phi}^{2} \right), \quad \textup{ with } \quad \tilde{\phi} = M\phi,
\end{equation}
and where $r$ and $\rho$ are related by

\begin{equation}\label{rho_r}
r^{2} = l^{2}M\csc^{2}{\rho}.    
\end{equation}

Now in order to draw the Penrose diagram,  by using (\ref{rho_r}), we can see that for the range of $\rho$, i.e., $\rho\in (-\pi,0) \cup (0,\pi)$, the regions that define the maximal extension of the spacetime correspond to    

\begin{eqnarray}
&& \textup{Wormhole throat: } \quad  \left(  r = r_{0} = \sqrt{l^{2} M} \right) \equiv \left( \rho = \pm\frac{\pi}{2} \right). \\
&& \textup{Asymptotic AdS regions: } \quad \left( r \sim \infty \right) \equiv \left( \rho = 0, \pm\pi \right).
\end{eqnarray}
\begin{center}
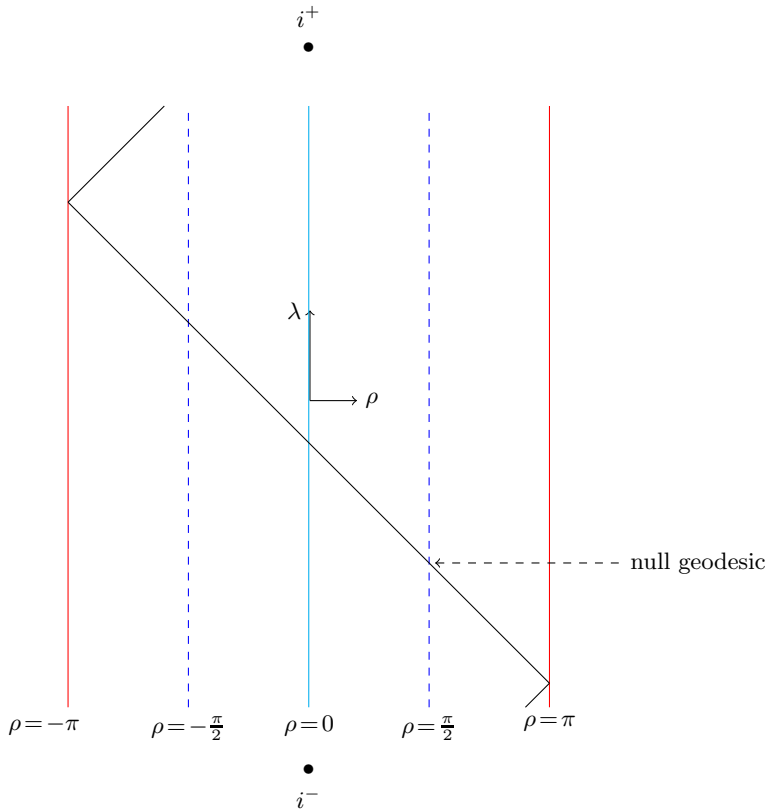
\begin{figure}[htb!]
\begin{tikzpicture}[scale=1.6]
\draw[->] (0.01,0.05) -- (0.4,0.05) node[right] {$\rho$}; 
\draw[->] (0.01,0.05) -- (0.01,0.8) node[left] {$\lambda$};
\draw[-, dashed, color=blue] (1,-2.5) -- (1,2.5);
\draw[-, dashed, color=blue] (-1,-2.5) -- (-1,2.5);
\draw[-, color=red] (2,-2.5) -- (2,2.5);
\draw[-, color=red] (-2,-2.5) -- (-2,2.5);
\draw[-,  color=cyan] (0,-2.5) -- (0,2.5);
\draw (-1.0,-2.5) node[below] {{\small $\rho\!=\! -\frac{\pi}{2}$}};
\draw (1,-2.5) node[below] {{\small $\rho\!=\! \frac{\pi}{2}$}};
\draw (-2.2,-2.5) node[below] {{\small $\rho\!=\! -\pi$}};
\draw (2,-2.5) node[below] {{\small $\rho\!=\! \pi$}};
\draw (0,-2.5) node[below] {{\small $\rho\!=\! 0$}};
\draw (0,-2.9) node[below] {$\bullet$};
\draw (0,3.1) node[below] {$\bullet$};
\draw (0,-3.1) node[below] {$ i^{-}$};
\draw (0,3.4) node[below] {$i^{+}$};
\draw[smooth, domain = -2:2, color=black]
plot (\x,{ -\x - 0.3 });
\draw[smooth, domain = -2.5:-2.3, color=black]
plot (\x + 4.3,{ \x   });
\draw[smooth, domain = 1.7:2.5, color=black]
plot (\x - 3.7,{ \x   });
\draw[<-, dashed, color=black] (1.05,-1.3) -- (2.6,-1.3) node[right]{null geodesic};
\end{tikzpicture}
\caption{Penrose diagram for the WH, metric (\ref{NewSolution2}): Continuous vertical lines show the spatial infinity, $\rho= - \pi, 0, \pi$; while the dashed ones represent  WH throats, $\rho= - \pi/2, \pi/2$. The black dots at the top and at the bottom, $ i^{+}$ and $ i^{-}$, denote the time infinity, future and past, respectively. 
Every light ray coming from infinity $\rho\sim\pm\pi$ will pass through the WH throat $\rho=\pm\pi/2$, and reach infinity $\rho\sim0$, and viceversa. In a similar way than for  the BTZ Penrose diagram, the WH  Penrose diagram can be embedded in the Einstein Universe.}
\label{figPfull}
\end{figure}
\end{center}
The Penrose diagram of the spacetime with metric (\ref{NewSolution2}) is shown in Fig. \ref{figPfull}. 
Finally,  symmetries  allow us to  consider the  Penrose diagram for the metric (\ref{NewSolution2}) being just half of the strip $\rho \in (-\pi,0) \cup (0, \pi)$, as shown in Fig. \ref{figPhal}. In a similar way than the BTZ Penrose diagram, the WH  Penrose diagram can be embedded into the Einstein Universe.
Thus, in Fig. \ref{figPhal}, every light ray coming from infinity $\rho \sim \pi$ will pass through the WH throat $\rho=\pi/2$, and reach infinity $\rho\sim0$, and viceversa. According to the  Penrose diagram,  anything that crosses the throat is not lost, but passes to the other part of the WH, in the extended manifold. 
\begin{center}
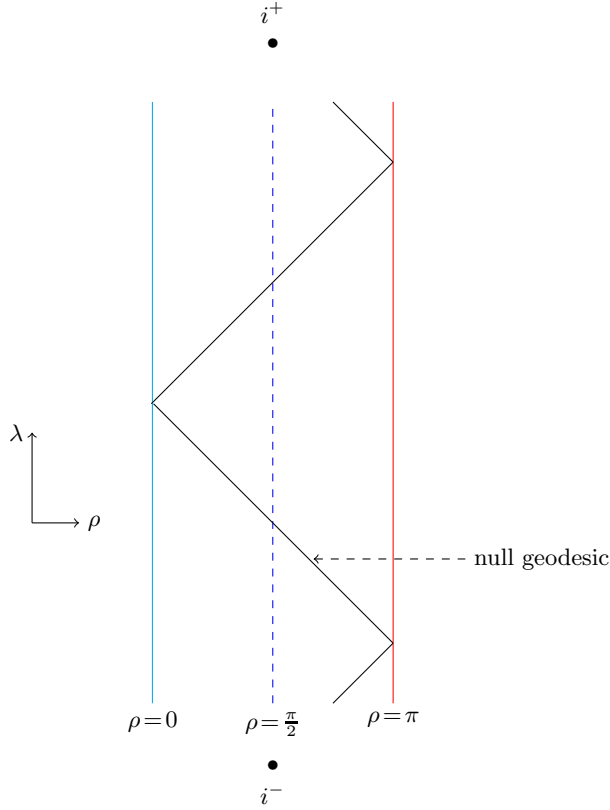
\begin{figure}[htb!]
\begin{tikzpicture}[scale=1.6]
\draw[->]  (-1,-1) -- (-0.61,-1) node[right] {$\rho$}; 
\draw[->] (-1,-1) -- (-1,-0.25) node[left] {$\lambda$};
\draw[-, dashed, color=blue] (1,-2.5) -- (1,2.5);
\draw[-, color=red] (2,-2.5) -- (2,2.5);
\draw[-, color=cyan] (0,-2.5) -- (0,2.5);
\draw (1,-2.5) node[below] {{\small $\rho\!=\! \frac{\pi}{2}$}};
\draw (2,-2.5) node[below] {{\small $\rho\!=\! \pi$}};
\draw (0,-2.5) node[below] {{\small $\rho\!=\! 0$}};
\draw (1,-2.9) node[below] {$\bullet$};
\draw (1,3.1) node[below] {$\bullet$};
\draw (1,-3.1) node[below] {$ i^{-}$};
\draw (1,3.4) node[below] {$i^{+}$};
\draw[smooth, domain = 2.5:2, color=black]
plot (-\x + 4,{ \x   });
\draw[smooth, domain = 0.0093206:2, color=black]
plot (\x,{ -\x   });
\draw[smooth, domain = -0.0093206:2, color=black]
plot (\x,{ \x   });
\draw[smooth, domain = -2.5:-2, color=black]
plot (\x + 4,{ \x   });
\draw[<-, dashed, color=black] (1.34,-1.3) -- (2.6,-1.3) node[right]{null geodesic};
\end{tikzpicture}
\caption{Penrose diagram for the WH, metric (\ref{NewSolution2}): Continuous vertical lines show the spatial infinity, while the dashed one represents the WH throat. Light rays coming from infinity $\rho \sim \pi$,  
will pass through the WH throat $\rho=\pi/2$, and then travel towards infinity $\rho \sim 0$.}
\label{figPhal} 
\end{figure}
\end{center}

\section{QNMs of a massive scalar test field in the WH spacetime} 

The QNMs encode the information on how  a perturbing field behaves in certain spacetime; they depend on the type of perturbation and on the geometry of the background system.  The QNMs of the BTZ-BH have been determined for a number of  perturbing fields, namely,  scalar, massive scalar, electromagnetic, etc \cite{Cardoso2001}, \cite{Birmingham2001}, \cite{Fernando2004}. In this section we address the perturbation of the previously introduced WH, (\ref{NewSolution2}), by a massive scalar  field  $\Psi(t,\vec{r})$. The effect is described by the solutions of the  KG equation,

\begin{equation}
(\nabla^{\alpha}\nabla_{\alpha} - \mu^{2})\Psi(t,\vec{r}) = 0,
\end{equation}
where $\mu$ is the mass of the scalar field; equivalently, the KG equation is,

\begin{equation}
\partial_{\alpha}\left( \sqrt{-g} g^{\alpha\beta}\partial_{\beta} \Psi(t,\vec{r}) \right) - \sqrt{-g}\mu^{2}\Psi(t,\vec{r}) = 0.    
\end{equation}   

The scalar field is suggested of the form
\begin{equation}\label{Psi}
\Psi(t,\vec{r}) = e^{-i\omega t}e^{i\ell\phi} R(r) ,   
\end{equation}
where $\omega$ is the frequency of the perturbation and  $\ell$ its azimuthal angular momentum. Substituting  (\ref{Psi}) into the KG equation, we arrive at  a second order equation for $R(r)$,

\begin{equation}\label{eqCR}
R'' +  \frac{2M l^2- 3r^2}{(Ml^2- r^2)r } R' - \left( \frac{ \omega^{2} - q_{\alpha} M^2 ( \ell^{2} + \mu^{2} r^2)}{q_{\alpha} M^2(M- {r^2}/{l^2}) r^2 }\right)R = 0.     
\end{equation}    
$R(r)$ is completely determined once the appropriate boundary conditions  are imposed. 
Since Eq. (\ref{eqCR}) diverges at the throat,  $r_{0}^2=-M/ \Lambda=M l^2$, it is useful to put the KG equation in terms of the tortoise  coordinate, $r_{\ast}$, defined in Eq. (\ref{tortuga}).
For the WH,  with $r_{0}$ being the throat, the boundary conditions for the QNMs consist in assuming purely ingoing waves at the throat of the WH,  $r=r_0$,  that in terms of the tortoise coordinate is $R(r) \sim e^{-i\omega r_{\ast}}$. While at infinity, $r \mapsto \infty$, the AdS boundary demands  the vanishing of the solution; i.e. the boundary conditions are  

\begin{eqnarray}
&&{ r \sim r_{0}  \Rightarrow R(r)  \sim e^{-i\omega r_{\ast}}, }\label{con_front_WH1}\\
&&{ r \sim \infty  \Rightarrow R(r) \sim 0. }\label{con_front_WH2}
\end{eqnarray}

\begin{figure}
\centering
\includegraphics[width=17.5cm,height=6cm]{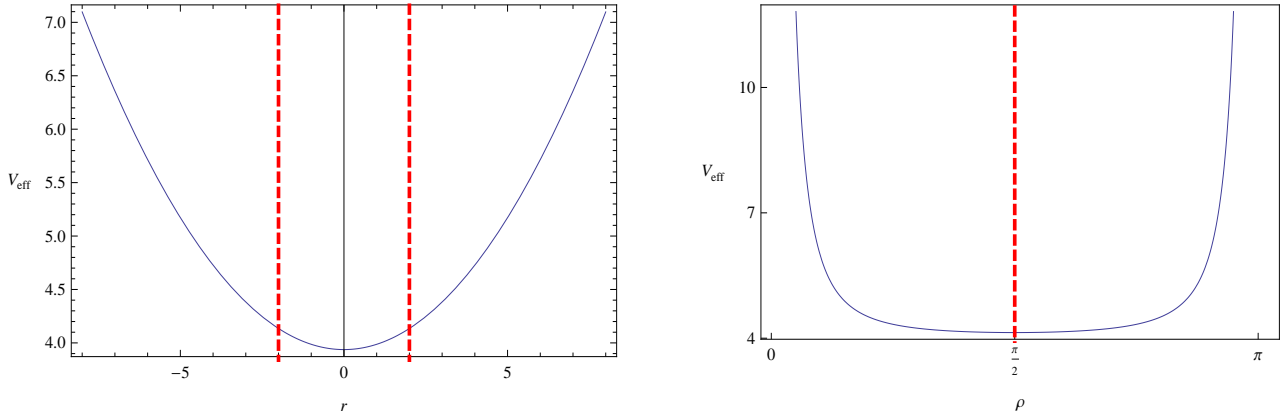}
\caption{\label{fig2} The effective potential as a function of the coordinate $r$ (left) and as a function of $\rho$ (right) are displayed; the dashed red vertical lines show the position of the throat, $r_0^2=M l^2$ or $\rho= \pi /2$. The parameters are fixed as $M=1$, $q_{\alpha}=0.5$ and the AdS parameter is $l=2$; while the field parameters are $\ell=4, \mu=0.1$. Note that the range of the coordinate $r$ is $r_{0} \le r < \infty$, i.e. to the right of the dashed red lines. }
\end{figure}

Let us return to the radial part of the KG Eq. (\ref{eqCR}). By transforming $R(r_{\ast}) = \psi (r_{\ast})/\sqrt{r}$ (considering $r$ as a function of $r_{\ast}$), we arrive at

\begin{equation}\label{Schr}
\ddot{  \psi  } (r_{\ast}) +  \left[ \omega^{2} -  V_{\rm eff}(r_{\ast})  \right]\psi (r_{\ast}) = 0,
\end{equation}
where $\dot{f}= df/dr_{\ast}$; while the effective potential $V_{\rm eff}$  in the Schr\"odinger-like equation  (\ref{Schr}), is

\begin{equation}
 V_{\rm eff}(r) =  q_{\alpha}^{2} M^{2} \left[ \left( \mu^{2}  + \frac{3}{4l^2} \right) r^{2} + \ell^{2} - \frac{M}{4} \right],  \quad\textup{ for }\quad r\in[r_{0},\infty) 
\end{equation}
that we identify as the potential of a displaced  harmonic oscillator, with frequency $\omega^2= q_{\alpha}^{2} M^{2} (\mu^{2}  + \frac{3}{4 l^2})$. Note that the displacement is proportional to the angular momentum of the scalar field, $ \ell^2$. The effective potential can be written  in terms of the tortoise coordinate as 

\begin{equation}
 V_{\rm eff}(r_{\ast}) =  q_{\alpha}^{2} M^{3} \left\{ \left(\mu^{2} l^2  + \frac{3}{4}\right) \csc^2\!\left( \sqrt{q_{\alpha}^{2} M^{3}} r_{\ast}\right) + \frac{\ell^{2}}{M} - \frac{1}{4} \right\}, \quad\textup{ for }\quad  \frac{\pi}{2 \sqrt{ q_{\alpha}^{2} M^{3} } }  \leq r_{\ast} < \frac{\pi}{   \sqrt{ q_{\alpha}^{2} M^{3} } }.
\end{equation}
The coordinate $\rho$, that was introduced in section \ref{Max_Ext}, as $r^{2} = l^{2}M\csc^{2}{\rho}$ [see  Eq. (\ref{rho_r})], in contrast to the coordinates $r$ and $r_{\ast}$,  covers both sides of the WH spacetime, side I:  $\rho\in(0 , \pi/2]$, and side II: $\rho\in[\pi/2 , \pi)$, connected by the WH throat located at $\rho = \frac{\pi}{2}$. The effective potential in terms of $\rho$ is

\begin{equation}\label{Vrho}
 V_{\rm eff}(\rho) =  q_{\alpha}^{2} M^{3} \left\{ \left(\mu^{2} l^2  + \frac{3}{4}\right) \csc^2\!(\rho) + \frac{\ell^{2}}{M} - \frac{1}{4} \right\}, \quad\textup{ for }\quad   \rho \in (0 , \pi). 
\end{equation}
The effective potential is depicted in Fig. \ref{fig2}, both, as a function of $r$ and of $\rho$.  It diverges at infinity, being, as a function of $r$, a confining harmonic oscillator-type potential; while as a function of $\rho$ it is a potential of the Rosen-Morse type \cite{Fdez2001}.

\subsection{The solution for the QNMs of the WH}

In general, the spectrum of the Schr\"odinger-like operator $\hat{H} = -\frac{d^{2}}{dr^{2}_{\ast}} + V_{\rm eff}(r_{\ast})$, can be decomposed into three parts: point spectrum (often called discrete spectrum); continuous spectrum and residual spectrum. In the case of our interest it turns out that the QNMs correspond to the point spectrum and it could be foreseen from the shape of the effective potential. 

In  \cite{Weyl}  H. Weyl showed that if $V(x)$ is a real valued continuous function on the real line $\mathcal{R}=(-L,L)$, $L\in\mathbb{R}$, and such that $\lim_{|x|\rightarrow L} V(x)\rightarrow \infty$, and such that $V(x)$ is monotonic in $|x|\in(-L,L)$, then the unbounded operator $-d^{2}/dx^{2} + V(x)$, acting on $\mathbf{L}^{2}(\mathcal{R})$, has pure point spectrum. Moreover, since $V(x)$ has the structure of an infinite well, it implies that all the eigenvalues of the operator $\hat{H} = -d^{2}/dx^{2} + V(x)$ will be real numbers, necessarily. 
Subsequently, in \cite{Titchmarsh}, this result was extended for the case in which $V(x)$ is not necessarily monotonic in $|x|\in(-L,L)$.

Specifically in our case, the potential $V_{\rm eff}(\rho)$, in Eq. (\ref{Vrho}), is a real valued continuous function in $(0 , \pi)$, and since $\lim_{|\rho|\rightarrow \pi} V_{\rm eff}(\rho) = \lim_{|\rho|\rightarrow 0} V_{\rm eff}(\rho) \rightarrow \infty$, then according to  \cite{Weyl,Titchmarsh}, the  the Schr\"odinger-like operator has a pure point spectrum. Thus we can conclude that the QNMs of the scalar field in the wormhole background (\ref{NewSolution2}) are purely real; i.e., these QNMs are in fact normal modes (NMs) of oscillations.  In agreement with the previous argument, the general solution of Eq. (\ref{Schr}) with $V_{\rm eff}(r_{\ast})$ in Eq.  (\ref{Vrho}), is given by

\begin{equation}\label{psi_Gen}
\psi(\rho) = B_{1} P^{Z}_{V}\left( \sqrt{  1 - \csc^{2}{\rho}   } \right)  +  B_{2}   Q^{Z}_{V}\left( \sqrt{ 1 - \csc^{2}{\rho} } \right) = B_{1} P^{Z}_{V}\left( i\cot{\rho}  \right)  +  B_{2}   Q^{Z}_{V}\left( i\cot{\rho} \right) ,       
\end{equation}
where $B_{1}$  and  $B_{2}$  are integration constants,  $P^{Z}_{V}(x)$  are the associated Legendre functions of the first kind, and  $Q^{Z}_{V}(x)$ are the  associated Legendre functions of the second kind; while the parameters  $V$  and  $Z$ are given, respectively, by

\begin{equation}\label{parVZ}
V = \sqrt{1 + \mu^{2}l^{2} }  - \frac{1}{2},  \quad  Z = \frac{  i}{\sqrt{M}} \sqrt{ \ell^{2} - \frac{M}{4} - \frac{\omega^{2}}{q_{\alpha}^{2}M^{2}} }.     
\end{equation}

\begin{figure}
\centering
\includegraphics[width=17.5cm,height=6cm]{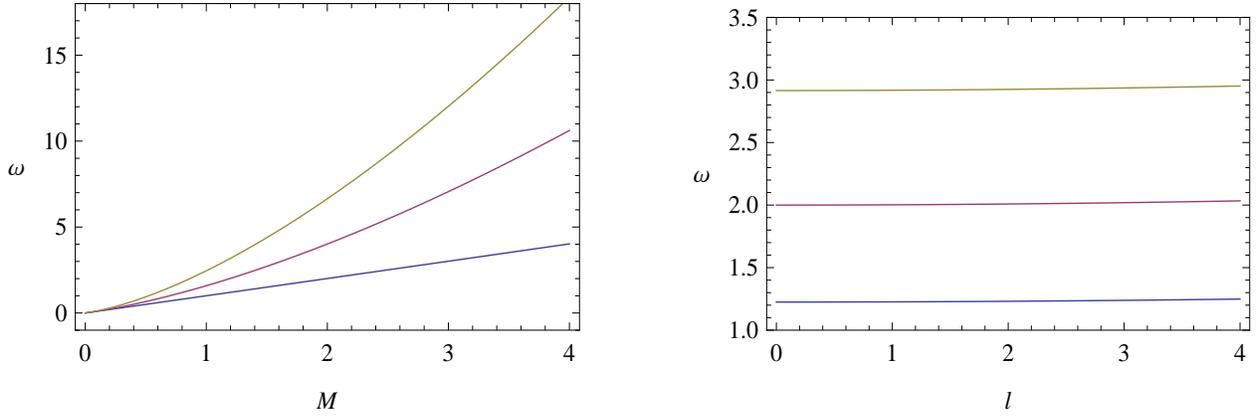}
\caption{\label{fig3} The quasinormal frequencies $\omega$ are shown, to the left as a function of the WH mass $M$ and to the right as a function of the AdS parameter $l$. The AdS parameter is $l=0.3$ in the plot to the left; $M=1$ in the graphic to the right.
The rest of parameters are fixed as $q_{\alpha}=0.5$;$\ell =2; \mu=0.1$. The QNMs are shown for  $n=0$, $n=1$ and $n=2$ in order from bottom to top.}
\end{figure}

For the sake of fluency in the text we skip the details on imposing the boundary conditions,  (\ref{con_front_WH1}) and (\ref{con_front_WH2}), and include them in the Appendix.
The boundary conditions for the QNMs consist in assuming purely ingoing waves at the throat of the WH,  $r=r_0$, that in terms of the tortoise coordinate are $R(r) \sim e^{-i\omega r_{\ast}} $.  While related to the AdS asymptotics it shall be required  the vanishing of the solution at  infinity, $r \mapsto \infty$; or in terms of $\rho$, $\rho \mapsto 0$ or $\rho \mapsto \pi$. These conditions imply restrictions in the values of the arguments of the Gamma functions related to the hypergeometric functions. Joining both conditions we arrive at the following restrictions for the WH parameters $(M, \Lambda, q_{\alpha})$ that combined with restrictions on the parameters of the perturbing field $(\ell, \mu, \omega)$ amount to

\begin{eqnarray}
&& 1 - Z + V = -2n, \quad  n\in \mathbb{N} + \{0\}, \label{ZV_set}\\
&&  \frac{1}{2} - i M^{-3/2}\sqrt{ \ell^{2}M^{2} - \frac{M^{3}}{4} - \left(\frac{\omega}{q_{\alpha}}\right)^{2}   }   + \sqrt{1 + \mu^{2}l^{2} }    = -2n, \\
&&\Rightarrow \omega^{2} = q_{\alpha}^{2} M^{3} \left[ \left(2n + \frac{1}{2} + \sqrt{ \mu^{2}l^2 + 1} \right)^2  + \frac{\ell^{2}}{M} - \frac{1}{4} \right]\label{wmodos}.
\end{eqnarray}
Besides, the  condition (\ref{ZV_set})  when applied to
the solution (\ref{psi_Gen}) it renders that the second term becomes a multiple of the first one, and then the solution (\ref{psi_Gen}),  in terms of the hypergeometric function, takes the form 

\begin{equation}\label{R_I_hypgeom}
\psi(\rho) =  \tilde{B}_{1} \left( \frac{ i\cot{\rho} + 1 }{ i\cot{\rho} - 1 } \right)^{ \frac{Z}{2} } \!\! \quad  _{2}\!\tilde{F}_{1}\!\!\left( -V, V + 1; 1 - Z; \frac{ 1 - i\cot{\rho} }{ 2  } \right),
\end{equation}
being $\tilde{B}_{1}$ a constant, while  $_{2}\!\tilde{F}_{1}\!(a,b;c;x)$ is a regularized  Gauss (or ordinary) hypergeometric function, related to the Gauss hypergeometric function $_{2}\!F_{1}\!(a,b;c;x)$ through $ _{2}\!\tilde{F}_{1}\!(a,b;c;x) =$ $_{2}\!F_{1}\!(a,b;c;x)/\Gamma(c)$, see Appendix for details.

Clearly from the expression for  $\omega^2$, Eq. (\ref{wmodos}), we see that it is real and always positive. It may be a surprise that $\omega$ is not complex, as corresponds to an open system, as deceptively may appear a WH. However a clue that this is not the case (WH as an open system) came from the form of the effective potential. The throat is not similar to the event horizon in a BH, where amounts of fields are lost once penetrating the horizon. In the case of a WH it is supposed that the waves that penetrate the throat are passing to the continuation of the WH. This image is in agreement to the Kruskal-Szekeres and Penrose diagrams. Moreover the fact that the frequency of the QNMs has not an imaginary part, tells us that the system will remain the same, i.e. massive scalar field solutions are stable.
In Fig. \ref{fig3} are plotted the frequencies as a function of the mass $M$ and of the AdS parameter $l$, for $n=0,1,2$; the tendency is of growing $\omega$ as $M$ increases, we can deduce a similar behavior for the variation of the electromagnetic parameter $q_{\alpha}$, and  a very slow increase of $\omega$ when $l$ increases.
In here we just note that for the hypergeometric function, if the first or the second argument is a non-positive integer, then the function reduces to a polynomial. In our problem it  is possible to impose that condition, by making $V=n$, being $n$ an integer; however, this is not our aim and we will not go further in this direction.

A remarkable particular case is $\ell = \pm\frac{\sqrt{M}}{2}$, 

\begin{equation}
\pm\omega =  \left(n + \frac{1}{2} \right)\omega_{0} +  \left(  \sqrt{  \mu^{2} l^2  + 1  } - \frac{1}{2} \right) \frac{ \omega_{0}}{2} , \textup{ with } \omega_{0} = 2|q_{\alpha}|\sqrt{M^{3}}.
\label{OSC_UEF}
\end{equation}

This spectrum resembles the corresponding to a quantum harmonic oscillator under the influence of an electric field  $\mathcal{E}$, 
$E_{n} = \left( n + \frac{1}{2} \right)\hbar \omega - \frac{q^{2}\mathcal{E}^{2}}{2m\omega^{2}}$,  with a frequency given by $\omega = \sqrt{\frac{k}{m}}$.
i. e. for this particular frequency, $\omega_{0} = 2\sqrt{M^{3} q_{\alpha}^2}$, the massive scalar field, confined by the WH-AdS spacetime,  will oscillate harmonically.  

\section{Final Remarks}

We have determined in exact form the QNMs of a massive scalar field in the background of a charged, static, cyclic symmetric (2+1)-dimensional traversable wormhole, determining that the characteristic frequencies are real and discrete (point spectrum), showing that as far as the test scalar field is concerned, the potential is a confining one.
The WH Penrose diagram agrees with this interpretation, since light trajectories passing through the WH throat arrive to the extended manifold, i.e. the other side of the WH.

Since there are no propagating degrees of freedom in the purely (2+1)-dimensional gravity,  it is important to couple (2+1)-gravity with other fields as well as probe (2+1)- systems with test fields such as scalar fields.  The BTZ-black hole has been of great relevance providing a mathematical model of a holographic manifold. Then (2+1)-systems in which quasinormal modes are exactly calculated, are encouraging examples for trying to go all the way through and find the correspondence with a holography theory \cite{Bertola2000}.
Holographic principle, roughly speaking, consists in finding a lower-dimensional dual field theory that contains the same information as gravity. In the system worked out in the present paper, we consider two fields, an electromagnetic field, characterized by a gauge, $A^{\mu}$ that could be the starting point to try a quantization scheme. We also show that the KG equation for a massive scalar field can be exactly solvable, providing then a scalar field, that could be used in searching for a correspondence in the AdS boundary. In other words, the system worked out here stimulates to explore the possibility of obtaining the conformal field associated to this  AdS-WH  solution in the  bulk, according to the  AdS/CFT correspondence.

\vspace{0.5cm}
\textbf{Acknowledgments}: N. B. and P. C. acknowledges partial financial support from CONACYT-Mexico through the project No. 284489. P. C. and L. O. thank Cinvestav for hospitality.

\section*{Appendix}

\appendix

In this Appendix, we present the details in setting the boundary conditions for the massive scalar field in the WH spacetime with metric (\ref{NewSolution2}).

\subsection{Boundary conditions at the throat $r=r_{0}$, and at infinity $r\sim\infty$ } 

Very close to the throat  $r \sim r_{0} = \sqrt{ -\frac{ M }{\Lambda} } = \sqrt{ l^{2}M }$  we shall require that $R(r_{\ast}) \sim e^{-i\omega r_{\ast}} $. For simplicity, the description will be presented in terms of the tortoise coordinate, $r_{\ast}$. Then, the asymptotic form of $e^{-i\omega r_{\ast}}$ near the throat $r_{\ast}  \sim \frac{ \pi }{ 2\sqrt{ q_{\alpha}^{2} M^{3} } }$, is given by 

\begin{equation}
 r_{\ast}  \sim \frac{ \pi }{ 2\sqrt{ q_{\alpha}^{2} M^{3} } } \Rightarrow e^{-i\omega r_{\ast}} \sim \left(e^{i\pi}\right)^{- \frac{ \omega }{ 2\sqrt{ q_{\alpha}^{2} M^{3} } } } = \left(  -1 \right)^{  -\frac{\omega}{ 2\sqrt{ q_{\alpha}^{2} M^{3} } } }  
\end{equation}
In such a way that  the condition  (\ref{con_front_WH1}) goes like

\begin{equation}\label{R_asy_r0}
r_{\ast}  \sim \frac{ \pi }{ 2\sqrt{ q_{\alpha}^{2} M^{3} } }  \Rightarrow R(r_{\ast}) = \frac{\psi(r_{\ast})}{ r^{\frac{1}{2}}\!(r_{\ast}) } 
\sim  
\left(  -1 \right)^{  -\frac{\omega}{ 2\sqrt{ q_{\alpha}^{2} M^{3} } } } \sim  {\rm constant}  \equiv \mathbb{C}-\{0\}.    
\end{equation}
To implement this asymptotic behavior in the solution $R(r_{\ast})=\psi(r_{\ast})/r^{\frac{1}{2}}\!(r_{\ast})$, with $\psi$ given in Eq. (\ref{psi_Gen}), with $\rho$ and $r_{\ast}$ related by $\rho = \sqrt{q_{\alpha}^{2} M^{3}} r_{\ast}$, recalling that the ranges are different, we shall write it as

\begin{equation}\label{R_1_y_R_2}
R(r_{\ast}) =   \frac{ B_{1} }{ r^{\frac{1}{2}}\!(r_{\ast}) } P^{Z}_{V}\!\!\left( i\cot(\sqrt{q_{\alpha}^{2} M^{3}} r_{\ast}) \right)  + \frac{ B_{2} }{ r^{\frac{1}{2}}\!(r_{\ast}) }  Q^{Z}_{V}\!\!\left( i \cot( \sqrt{q_{\alpha}^{2} M^{3}} r_{\ast}) \right) = B_{1} R_{I}(r_{\ast}) + B_{2} R_{II}(r_{\ast}), 
\end{equation}
the  two terms,  $R_{I}(r_{\ast}) = \frac{ 1 }{ r^{\frac{1}{2}}\!(r_{\ast}) } P^{Z}_{V}\!\!\left( i \cot( \sqrt{q_{\alpha}^{2} M^{3}} r_{\ast}) \right)$ and $R_{II}(r_{\ast}) = \frac{ 1 }{ r^{\frac{1}{2}}\!(r_{\ast}) } Q^{Z}_{V}\!\!\left(i \cot( \sqrt{q_{\alpha}^{2} M^{3}} r_{\ast}) \right)$ shall be analyzed separately. Moreover,  in terms of the hypergeometric functions, the quantities $R_{I}(r_{\ast})$ and $R_{II}(r_{\ast})$ become 
\begin{equation}\label{R_I_hypgeom}
R_{I}(r_{\ast}) =   \frac{ 1}{ r^{\frac{1}{2}}\!(r_{\ast})  }\left( \frac{  i \cot( \sqrt{q_{\alpha}^{2} M^{3}} r_{\ast}) + 1 }{ i \cot(\sqrt{q_{\alpha}^{2} M^{3}} r_{\ast}) - 1 } \right)^{  \frac{  i}{2\sqrt{M}} \sqrt{ \ell^{2} - \frac{M}{4} - \frac{\omega^{2}}{q_{\alpha}^{2}M^{2}} } }  \!\!\!\!\! \quad  _{2}\!\tilde{F}_{1}\!\!\left( -V, V + 1; 1 - Z; \frac{ 1 - i\cot( \sqrt{q_{\alpha}^{2} M^{3}} r_{\ast})  }{ 2 } \right),
\end{equation}
The behavior of $R(r_{\ast})$  at the wormhole throat $r_{\ast}  \sim \frac{ \pi }{ 2\sqrt{ q_{\alpha}^{2} M^{3} } }$   
(in this neighborhood $l^{2}M - r^{2} \sim 0$) is

\begin{equation}\label{R_I_asy1}
R_{I}( r_{\ast} ) \sim  \frac{ 1 }{ \sqrt{ \sqrt{ l^{2}M  } } } \left( \frac{  i \cot( \sqrt{q_{\alpha}^{2} M^{3}} r_{\ast}) + 1 }{ i \cot(\sqrt{q_{\alpha}^{2} M^{3}} r_{\ast}) - 1 } \right)^{  \frac{  i}{2\sqrt{M}} \sqrt{ \ell^{2} - \frac{M}{4} - \frac{\omega^{2}}{q_{\alpha}^{2}M^{2}} } }  \!\!\!\!\!\quad  _{2}\!\tilde{F}_{1}\!\!\left( -V, V + 1; 1 - Z; \frac{1}{2} \right). 
\end{equation}
Now, using the Bailey's summation theorem

\begin{equation}\label{Bailey}
_{2}\!F_{1}\!\!\left( a, 1 - a; c ;\frac{1}{2} \right) =  \frac{\Gamma(\frac{c}{2})\Gamma(\frac{1+c}{2})}{\Gamma(\frac{c+a}{2})\Gamma(\frac{1+c-a}{2})},
\end{equation}
the Eq. (\ref{R_I_asy1}) takes the form

\begin{equation}\label{RI_g}
R_{I}(r_{\ast}) \sim   \frac{ 1 }{ \sqrt{ \sqrt{ l^{2}M  } } } \left( \frac{  i \cot( \sqrt{q_{\alpha}^{2} M^{3}} r_{\ast}) + 1 }{ i \cot(\sqrt{q_{\alpha}^{2} M^{3}} r_{\ast}) - 1 } \right)^{  \frac{  i}{2\sqrt{M}} \sqrt{ \ell^{2} - \frac{M}{4} - \frac{\omega^{2}}{q_{\alpha}^{2}M^{2}} } } \frac{\Gamma(\frac{1-Z}{2})\Gamma(\frac{2-Z}{2})}{\Gamma(\frac{1-V-Z}{2})\Gamma(\frac{2+V-Z}{2})} \sim R_{\ast}\frac{\Gamma(\frac{1-Z}{2})\Gamma(\frac{2-Z}{2})}{\Gamma(\frac{1-V-Z}{2})\Gamma(\frac{2+V-Z}{2})}, 
\end{equation}
where $R_{\ast}\in\mathbb{C}-\{0\}$ is constant. While in order that $R_{I}(r_{\ast})$ behaves properly, the factor 
$\Gamma(\frac{1-Z}{2})\Gamma(\frac{2-Z}{2})/ (\Gamma(\frac{1-V-Z}{2})\Gamma(\frac{2+V-Z}{2}))$
should be finite and non-vanishing; this can be accomplished if   $\frac{1-Z}{2}$, $\frac{2-Z}{2}$, $\frac{1-V-Z}{2}$ and $\frac{2+V-Z}{2} = \frac{1}{2} + \frac{1 + V-Z}{2}$, are not in the set $\{$ 0, -1, -2, -3, .... -$n$,... $\}$ with $n \in \mathbb{N};$ consequently,  $1+V-Z$ $\neq$ $-1$, $-3$, $-5$, $-7$, .. $-(2n+1)$.  With this restriction we get to
 
\begin{equation}\label{RI_clo_g}
r_{\ast} \sim \frac{ \pi }{ 2\sqrt{ q_{\alpha}^{2} M^{3} } } \Rightarrow R_{I}(r_{\ast}) \sim  {\rm constant}  \in \mathbb{C} - \{0\}, 
\end{equation}
Having then accomplished that for $ r\sim r_{0} = \sqrt{l^{2}M} \Rightarrow  R_{I}(r_{\ast}) \sim e^{-i\omega r_{\ast}}$ [condition (\ref{con_front_WH1})]. 

In order to get the function that describes the asymptotic behavior of the second term in $R(r_{\ast})$ as $r\sim r_{0} = \sqrt{ l^{2}M }$,

$R_{II}(r_{\ast}) = \frac{ 1 }{ r^{ \frac{1}{2} }(r_{\ast}) } Q^{Z}_{V}\!\!\left( i\cot(\sqrt{q_{\alpha}^{2} M^{3}} r_{\ast}) \right)$, we write it in terms of the hypergeometric functions,

\begin{eqnarray}\label{R_II_exact}
R_{II}( r_{\ast} ) &=& \frac{\pi}{2} R_{I}( r_{\ast} ) \nonumber \\
&-&   \frac{\Gamma(Z+V+1)}{\Gamma(-Z+V+1)} \frac{\pi\csc{(Z\pi)}}{ 2 r^{ \frac{1}{2} }\!(r_{\ast}) }\left( \frac{ i\cot(\sqrt{q_{\alpha}^{2} M^{3}} r_{\ast}) - 1 }{ i\cot(\sqrt{q_{\alpha}^{2} M^{3}} r_{\ast}) + 1 }\right)^{\!  \frac{ i }{ 2\sqrt{M} } \sqrt{ \ell^{2} - \frac{M}{4} - \frac{\omega^{2}}{q_{\alpha}^{2}M^{2}} } }  \!\!\!\!\!\!\!\quad  _{2}\!\tilde{F}_{1}\!\!\left( -V, V + 1; 1 + Z; \frac{ 1 - i\cot(\sqrt{q_{\alpha}^{2} M^{3}} r_{\ast}) }{ 2 } \right).\nonumber \\     
\end{eqnarray}
Now we can analyze the behavior of $R_{II}$  as $r_{\ast} \sim \frac{ \pi }{ 2\sqrt{ q_{\alpha}^{2} M^{3}}}$, 

\begin{eqnarray}
R_{II}(r_{\ast}) &\sim&  \frac{\pi}{2} R_{I}(r_{\ast}) - \frac{\Gamma(Z+V+1)}{\Gamma(-Z+V+1)} \frac{\pi\csc{(Z\pi)}}{2\sqrt{\sqrt{ l^{2}M }}}\left( \frac{ i\cot(\sqrt{q_{\alpha}^{2} M^{3}} r_{\ast}) - 1 }{ i\cot(\sqrt{q_{\alpha}^{2} M^{3}} r_{\ast}) + 1 } \right)^{  \frac{  i}{2\sqrt{M}} \sqrt{ \ell^{2} - \frac{M}{4} - \frac{\omega^{2}}{q_{\alpha}^{2}M^{2}} } } \!\!\!\!\!\quad  _{2}\!\tilde{F}_{1}\!\!\left( -V, V + 1; 1 + Z; \frac{1}{2} \right), \nonumber \\  
&\sim&  \frac{\pi}{2} R_{I}( r_{\ast} ) - \frac{ \Gamma(Z+V+1) }{ \Gamma(-Z+V+1) } \frac{ \pi\csc{(Z\pi)} }{ 2\sqrt{ \sqrt{ l^{2}M } } }   \left( \frac{ i\cot(\sqrt{q_{\alpha}^{2} M^{3}} r_{\ast}) - 1 }{ i\cot(\sqrt{q_{\alpha}^{2} M^{3}} r_{\ast}) + 1 } \right)^{  \frac{  i}{2\sqrt{M}} \sqrt{ \ell^{2} - \frac{M}{4} - \frac{\omega^{2}}{q_{\alpha}^{2}M^{2}} } }  \frac{\Gamma(\frac{1+Z}{2})\Gamma(\frac{2+Z}{2})}{\Gamma(\frac{1-V+Z}{2})\Gamma(\frac{2+V+Z}{2})}\\
&\sim&  \frac{\pi}{2} R_{I}( r_{\ast} ) -   \frac{\Gamma(Z+V+1)}{\Gamma(-Z+V+1)} \frac{\pi \csc{(Z\pi)}}{2 \sqrt{\sqrt{ l^{2}M }} }( -1)^{  \frac{  i}{2\sqrt{M}} \sqrt{ \ell^{2} - \frac{M}{4} - \frac{\omega^{2}}{q_{\alpha}^{2}M^{2}} } }  
\frac{\Gamma(\frac{1+Z}{2})\Gamma(\frac{2+Z}{2})}{\Gamma(\frac{1-V+Z}{2})\Gamma(\frac{2+V+Z}{2})}.
\end{eqnarray}

Therefore we have determined the asymptotic behavior of $R_{II}( r_{\ast} )$ at the throat. We considered that as  $r_{\ast} \sim \frac{ \pi }{ 2\sqrt{ q_{\alpha}^{2} M^{3} } }$ i.e., ($r \sim r_{0}$), then $ l^{2}M - r^{2}\sim 0$, and we also used Eq. (\ref{Bailey}).
\subsection{Boundary condition at  infinity, $(r\sim\infty)$. } 

In what follows we shall impose the second boundary condition at infinity,  $r\sim\infty$; in terms of the tortoise coordinate is equivalent to $r_{\ast} \sim \frac{ \pi }{ \sqrt{ q_{\alpha}^{2} M^{3} } }$, then  $R(r_{\ast})\rightarrow0$ [condition \ref{con_front_WH2}]. It will be done separately for  $ R_{I}(r_{\ast})$ and $R_{II}(r_{\ast})$. 

It shall be considered first the term  $R_{I}(r_{\ast})$ written in terms of the hypergeometric function, Eq. (\ref{R_I_hypgeom}).  Since the last argument 
$x=(1 - i\cot(\sqrt{q_{\alpha}^{2} M^{3}} r_{\ast}) )/2$ of the hypergeometric function diverges when  $r_{\ast}\sim\frac{ \pi }{ \sqrt{ q_{\alpha}^{2} M^{3} } }$ the following identity can be used,

\begin{eqnarray}
 _{2}\!F_{1}\!\!\left( a, b; c;  x \right) &=& \Gamma(c) \!\!\!\!\!\quad _{2}\!\tilde{F}_{1}\!\!\left( a, b; c;  x \right) =\frac{ \Gamma(c)\Gamma(b-a) }{ \Gamma(b)\Gamma(c-a) } (-x)^{-a}    \!\!\!\!\!\quad  _{2}\!F_{1}\!\!\left( a, a - c + 1; a - b + 1;  \frac{1}{x} \right) \nonumber \\
 &+&  \frac{ \Gamma(c)\Gamma(a-b) }{ \Gamma(a)\Gamma(c-b) } (-x)^{-b} \!\!\!\!\!\quad  _{2}\!F_{1}\!\!\left( b, b - c + 1; b - a + 1;  \frac{1}{x} \right),  
\end{eqnarray}
that  allows us to write the asymptotic expression for  $ R_{I}(r_{\ast})$ as

\begin{eqnarray}
R_{I}(r_{\ast})  &\sim&   \frac{ \Gamma(2V+1)C_{1}  }{ \Gamma(V+1)\Gamma(1-Z+V) }  r^{V-\frac{1}{2}}(r_{\ast})  \!\!\!\!\!\quad  _{2}\!\tilde{F}_{1}\!\!\left( -V, Z-V; -2V; 0 \right)   \nonumber \\  
&& + \!\!\!\quad  
\frac{ \Gamma(-2V-1)C_{2} }{ \Gamma(-V)\Gamma(-Z-V) } r^{-V-\frac{3}{2}}\!(r_{\ast}) \!\!\!\!\!\quad  _{2}\!\tilde{F}_{1}\!\!\left( V+1, Z+V+1; 2V+2; 0 \right),      
\end{eqnarray}
where $C_{1}$ and  $C_{2}$ are complex constants. Using now that $_{2}\!F_{1}\!\!\left( a, b; c; 0 \right) = \Gamma(c) \!\!\!\!\!\quad _{2}\!\tilde{F}_{1}\!\!\left( a, b; c;  0 \right) =1/\Gamma(c)$, the previous equation can be written as

\begin{equation}\label{RI_infy}
R_{I}(r_{\ast})  \sim   \frac{ \Gamma(2V+1)C_{1}  }{ \Gamma(V+1)\Gamma(1-Z+V) } \frac{1}{ \Gamma^{2}(-2V) }   r^{ V-\frac{1}{2}}(r_{\ast})            
 + \!\!\!\quad  
\frac{ \Gamma(-2V-1)C_{2} }{ \Gamma(-V)\Gamma(-Z-V) }  \frac{1}{ \Gamma^{2}(2V+2) } r^{ -V-\frac{3}{2} }(r_{\ast}).
\end{equation}
On the other hand, given that $V = \sqrt{ 1 + \mu^{2}l^{2}   } - \frac{1}{2}$, and since $\mu$, $l$ $\in \mathbb{R}$ $\Rightarrow$ $\sqrt{ 1 + \mu^{2}l^{2}   }  > 1 $ $\Rightarrow$ $V > \frac{1}{2}$.
Then the behavior of  $R_{I}(r) $ goes like

\begin{equation}
\lim_{r\rightarrow\infty}  r^{ V-\frac{1}{2}}    \rightarrow \infty , \quad {\rm and}  \lim_{r\rightarrow\infty} r^{ -V-\frac{3}{2} }  \rightarrow 0.
\end{equation}

Therefore, the fulfilment of the proper behavior,  $R_{I}(r)  \sim 0$  when $r\rightarrow\infty$, and keeping the convergence of the second term in (\ref{RI_infy}),   imposes the condition that  $(-2V-1) \neq 0, -1, -2, -3,...-n,...$ with  $n\in \mathbb{N}$,  guaranteeing then that $\Gamma(-2V-1)$ be finite. 
In other words,  $(-2V-1) \neq 0, -1, -2, -3,...-n,... \Rightarrow (-2V) \neq 1, 0, -1, -2, -3,...-n,...$,  implying that $1 / \Gamma(-2V) \neq 0$.
Moreover,  $V+1 > 0$  implies that $ 1/\Gamma(V+1)  \neq 0$,   but the fulfilment of the boundary condition that $R(r)$ vanishes at infinity requires that 
$ 1/ \Gamma(1-Z+V) = 0$; this condition imposes that $(1-Z+V) = 0, -1, -2, -3,...-n,...$  with $n\in \mathbb{N}$. This guarantees the vanishing of the first term in (\ref{RI_infy}), accomplishing then the desired behavior at infinity.

We still have to consider the compatibility of the previously determined values of $V$ and $Z$  with the  fulfilment of  the first boundary condition, that at the throat  $R_{I}(r_{\ast})  \sim e^{-i\omega r_{\ast}}$. Eq. (\ref{RI_clo_g}) imposes that $(1-Z+V)\neq -1, -3, -5, -7,... -(2n+1),..$ with $n\in \mathbb{N}$.
Gathering the two conditions lead us to the  following 

\begin{equation}
\left\{ (1-Z+V) = -n : n\in \mathbb{N} \right\}  -  \left\{ (1-Z+V) = -(2n+1) : n\in \mathbb{N} \right\}  = \left\{ (1-Z+V) =  0, -2, -4, -6,...-2n,..\right\},   
\end{equation}
i.e.   $R_{I}(r_{\ast})$  has the asymptotic behaviors  (\ref{con_front_WH1})  and  (\ref{con_front_WH2}) provided

\begin{equation}\label{paramsbuen}
1 - Z + V = -2n, \textup{ with } n\in \mathbb{N} + \{0\}. 
\end{equation}

\subsection{ Behavior of   $R_{II}(r_{\ast})$ at infinity}

In Eq.  (\ref{R_II_exact}) was defined  $R_{II}(r_{\ast})$. Substituting the previously derived condition (\ref{paramsbuen}) into (\ref{R_II_exact}), leads to the vanishing of the second term of  $R_{II}(r_{\ast})$ since  $ 1/\Gamma( - Z + 1 + V) = 0$. This will occur whenever (i) $Z$ is not an integer, otherwise  $\csc{\!(Z\pi)}$ will diverge; and (ii)  $1+Z+V \neq - n$ with $n\in\mathbb{N} + 0$, otherwise $\Gamma(1+Z+V)$ diverges.

Summarizing, the fulfilment of the condition  (\ref{paramsbuen}), along with   $Z\neq \pm n$  and $1+Z+V \neq - n, \quad n\in\mathbb{N} + 0$, leads to the following simplifications, 

\begin{equation}
R_{II}(r_{\ast}) = \frac{\pi}{2} R_{I}(r_{\ast}) \Rightarrow R(r_{\ast}) = B_{1}R_{I}(r_{\ast}) + B_{2}R_{II}(r_{\ast}) = \tilde{B}_{1}R_{I}(r_{\ast}), \quad  \textup{ i.e., }  R(r_{\ast}) = \tilde{B}_{1} R_{I}(r_{\ast}), \quad  \forall r_{\ast}.
\end{equation}

Being then achieved the fulfilment of the two boundary conditions, at the throat and at infinity, for the solution $R(r_{\ast})$ for the QNMs of the scalar test field coming from the WH.


\end{document}